\documentstyle[preprint,eqsecnum,prd,aps]{revtex}
\input{epsf}
\begin{document} 
\renewcommand{\thefootnote}{\fnsymbol{footnote}}
\setcounter{equation}{0}
\newcommand{\beq}{\begin{equation}}
\newcommand{\eeq}{\end{equation}}
\newcommand{\beqa}{\begin{eqnarray}}
\newcommand{\eeqa}{\end{eqnarray}}
\input{epsf}
\pagestyle{plain}

\preprint{
\vbox{
\halign{&##\hfil\cr
	& ANL-HEP-PR-96-32\cr
        & MC-TH-96/15 \cr}}
}
\title{Prompt Photon Production in $\gamma \gamma$ Collisions
and the Gluon Content of the Photon}

\author{L. E. Gordon$^a$ and J. K. Storrow$^b$}
\address{
$^a$ High Energy Physics Division, Argonne National Laboratory,
	Argonne, IL 60439, USA \\ $^b$ 
Dept. of Theoretical Physics, University of Manchester,
 Manchester M13 9PL, England}

\maketitle
\begin{abstract} 
We calculate the cross section for inclusive prompt photon production 
in $\gamma\gamma$ collisions, i.e. the reaction $e^+e^-\to
\gamma \gamma \to \gamma X$, in next-to-leading order QCD. 
We show that at LEP2 energies this cross section is measurable and is 
sensitive to the gluon distribution of the photon, $g^{\gamma}$, which is 
currently very poorly constrained by data.
\end{abstract}
\vspace{0.2in}
\pacs{12.38.Bx, 13.65.+i, 12.38.Qk}

\narrowtext
\section{Introduction}

The quark parton distribution functions (pdfs) of various particles, 
such as the proton and photon, 
are mainly determined by structure function measurements. This is because 
in leading order (LO) the structure function $F_2(x,Q^2)$ is directly 
proportional to the quark and anti-quark distributions:
\begin{equation}
F_{2}(x,Q^2)=\sum^{N_f}_{i=1} x e^2_i\left(q_i(x,Q^2)+ {\overline
{q}}_i(x,Q^2)\right)
\end{equation}
However gluon distributions are poorly constrained by such analyses 
because of the relatively weak coupling between the Altarelli-Parisi equations 
for the singlet quark and gluon sectors and so less direct methods 
must be used. In the case of hadrons, in particular the proton, the 
momentum sum rule provided the initial evidence for the 
existence of gluons and remains a very powerful tool in constraining the 
gluon distribution, particularly when combined with theoretical constraints 
on the $x$-dependence of the distributions as provided by counting rules and
Regge theory. The application of perturbative QCD (PQCD) to phenomena 
at large momentum transfers provides invaluable further constraints on the 
pdfs. In particular, prompt photon production at large 
momentum transfer ($p_T$) is particularly useful, as it is very 
sensitive to the gluon distributions of the colliding particles as 
the dominant hard subprocess here is $gq\to \gamma q$. However, a 
meaningful analysis of existing data \cite{CDF,UA2} on ${\overline p} p$ 
collisions is highly
non-trivial: in particular, isolation criteria on the photon signal must
be imposed in order to reduce the background to the subprocess of 
interest produced by non-prompt photons from $\pi ^0$s produced at 
large $p_T$ and subsequently decaying into two photons and prompt 
photons produced by fragmentation from the final state partons
\cite{BQ,GGRV}.     

For the case of the photon, the situation is much worse than for the
proton and the pdfs relatively 
poorly known for various reasons. Firstly, the structure function data
on $F_2^{\gamma}(x,Q^2)$ are much less precise than that for protons 
and so the knowledge of the quark distributions is correpondingly worse. In 
addition, the gluon pdf, $g^{\gamma}$, is 
even less well determined. {\bf Most importantly, it is not constrained by a 
momentum sum rule }\cite{GS,JKSLUND}. Also 
jet studies in photon induced reactions are in 
their infancy. However jet cross-section measurements at TRISTAN
\cite{AMY,TOPAZ} have recently  established that
$g^{\gamma}\neq 0$, a result confirmed at HERA \cite{H1}.

The result of all this is that the available parametrizations of the photon 
have considerably different gluon distributions\cite{VOGTLUND,SSV}. The 
input gluon 
distributions in the evolution equations, whilst not completely arbitrary, 
are currently just theoretically motivated guesses. Further PQCD studies of
large $p_T$ jets in photoproduction at HERA and in $\gamma \gamma $ 
collisions with better data than are currently available 
will improve this situation but it seems 
worth investigating whether the production of prompt photons at large $p_T$ 
in photon initiated reactions can be used to extract information on 
$g^{\gamma}$. There have been several studies of prompt photon 
production at HERA \cite{ACFGP,GSG,GV}. The more realistic
analyses\cite{GSG,GV}, which work in the HERA lab frame with a spectrum 
of initial photon energies given by the equivalent photon 
approximation (EPA)\cite{EPA}, find slightly discouraging results in
the sense that the competing subprocesses are hard to disentangle and
the signal due specifically to  $g^{\gamma}$ difficult to isolate.

In this letter we investigate prompt photon production in $\gamma
\gamma$  collisions as a means of constraining $g^{\gamma}$. This process was 
first investigated by Drees and Godbole \cite{DG}: their work must
be regarded as a preliminary effort for several reasons. Firstly, it was  
a LO calculation of prompt photon plus opposite side jet.
The reason for the latter condition was to suppress the 
contribution of fragmentation processes by requiring a kinematic balance 
(in $p_T$) between the photon and opposite side jet: because of the 
difficulty in measuring the `true' $p_T$ of the
jet it is not clear how efficient this requirement would be. Also the 
photon pdfs they used have been superseded \cite{SSV}. 
In our study we consider the {\bf inclusive} prompt photon cross 
section including fragmentation contributions. We calculate in 
next-to-leading order (NLO) QCD using
up-to-date NLO photon pdfs and NLO photon fragmentation functions (FFs). 
 
\section{Basic Mechanisms}

To answer the question as to whether this cross section will be 
useful in constraining $g^{\gamma}$, there are various points we need to 
consider:
 
\begin{enumerate}
\item{Firstly in LO, the process is $O(\alpha_{em}^3/\alpha_s)$ where 
$\alpha_{em}$ and $\alpha_s$ are the electromagnetic and strong coupling 
respectively, so we need to know whether the cross section will be large 
enough to measure, at least at LEP2.}

\item{There are many subprocesses contributing to the cross section which we 
shall list below.  For the purpose of extracting $g^{\gamma}$, the most 
important subprocesses are the ones involving two resolved \cite{DGO} 
photons in the 
initial state. We need to know whether there are any accessible kinematic 
regions where these processes contribute significantly.} 

\item{As we shall soon see, there are many subprocesses where the prompt 
photon is produced via fragmentation off a final state parton. These 
involve the photon FFs,  which are not well known 
at present. Thus it is important that the contribution of 
these background processes is not very significant.}

\item{Finally, we need to determine how important the NLO corrections are in 
order to decide whether conclusions drawn from the LO study are valid when 
HO corrections are taken into account. The only way to determine this is to 
calculate the cross section fully in NLO.}
\end{enumerate}

\subsection{Contributing Subprocesses: LO}

We begin by discussing the different contributions in the LO case. We divide
the contribution into the seven types shown in fig.1(a)-(g) and classified 
in Table 1. They are classified according to the initial state, i.e.
whether one (1-res), both (2-res), or neither (D), of the initial 
state photons are resolved , the final state, i.e. whether the prompt 
photon is produced directly 
in the hard subprocess (NF) or by the subsequent fragmentation 
from one of the partons produced in the hard subprocess(F), and by 
the type of hard subprocess. The notation for resolved photons and 
photon fragmentation is described in fig.1(h). All of these
contributions in
fig.1(a-g) are $O(\alpha_{em}^3/\alpha_s)$ because both the photon pdfs and 
photon FFs are $O(\alpha_{em}/\alpha_s)$. To take an
example we consider the contribution shown in fig.1(c). This 
involves one photon pdf and one photon FF which when convoluted 
with the  subprocess cross section $\gamma q \to gq$ which is 
$O(\alpha_{em}\alpha_s)$ yields a contribution of
$O(\alpha_{em}^3/\alpha_s)$. 
The 2-res fragmentation processes generically shown in fig.1(e) contain 
many type of subprocess, namely: 
\begin{eqnarray}
q+ q &\rightarrow & q+ q \nonumber \\
q+ q' &\rightarrow & q +q' \nonumber \\
q+ {\overline q} &\rightarrow & q +{\overline q} \nonumber \\
q +{\overline q} &\rightarrow & q'+ {\overline q'} \nonumber \\
q +{\overline q} &\rightarrow & g+g \nonumber \\
g+ g &\rightarrow & q+ {\overline q} \nonumber \\
q+ g &\rightarrow & q+ g \nonumber \\
g+ g &\rightarrow & g+ g \nonumber \\
\end{eqnarray}
and we sum over them all. The total is still small.

Note that as regards a clean signal from the gluon content of the photon, 
not involving fragmentation, it is
mostly the NF 2-res process of fig.1(g) that is relevant, corresponding 
to the subprocess $gq\to \gamma q$. Fortunately it
is very significant, as we shall see.

\subsection{Contributing Subprocesses: NLO}

When it comes to the NLO calculation, then there are four types of 
corrections to the basic mechanisms:
\begin{enumerate}

\item{NLO corrections to photon pdfs,}

\item{NLO corrections to the photon FFs,}

\item{NLO corrections to the matrix elements for the 1-res and 2-res
processes,}

\item{NLO corrections to the direct contribution.}
\end{enumerate}  

For (1) and (2), we simply use pdfs and FFs valid in NLO, as they are 
available. For (3), the matrix 
elements for these are available \cite{AUR,GOR,AVER} and have previously
been used in calculations of prompt photon production at HERA \cite{GV}. 
Finally for (4), we need to evaluate the gluonic radiative corrections 
(both real and virtual) to the LO process of fig.1(a). Examples of
these are shown in figs. 2(a) and 2(b). At NLO the $O(\alpha_{em}^3)$
process $\gamma\gamma\rightarrow q\bar{q}\gamma$ also contributes. 
We did not need to recalculate the matrix elements for this contribution
since it can be simply obtained from the  real gluon  radiative process 
$\gamma\gamma\rightarrow q\bar{q}g$ \cite{GOR} (depicted in
fig.2(a) but before convolution with the fragmentation function) by adjusting 
couplings and color factors. 
All these NLO processes contribute up to $O(\alpha_{em}^3)$.

\section{Results}

For our calculations we use the EPA with the anti-tagging
angle set at 35 milliradians, relevant to the LEP2 detectors. For the 
photon pdfs we use those of refs \cite{GS,GSP1,GSP2,GRV}: for the 
FFs we use those of ref \cite{GRVF}. For the scale of the hard
scattering, $Q^2=(p_T^\gamma)^2$ is chosen throughout. In fig.3 we show the 
cross section vs $p_T^\gamma$ integrated over the rapidities in the 
range $-2\le y^\gamma \le 2$ at the $e^+e^-$ CMS energy $\sqrt{s}=180$ GeV. 
This rapidity cut is relevant for the LEP2 detectors in the sense that 
to distinguish between electrons and photons the track must pass through
the central tracking detector. In fig.3(a) we show the LO 
and NLO results using the new GS photon distributions and the NLO result
using the GRV distributions. The K-factor in not very different from $1$
even at low $p_T^\gamma $ values, indicating perturbative stability of the
result. If one compares the LO and NLO results using the GRV
distributions one finds significantly larger K-factors, as the LO GRV
distributions give very similar results to the LO GS distributions. The 
reason for this difference is not completely clear, but it could be due
to the fact that the LO and NLO GRV photon pdf parametrizations \cite{GRV} 
were fitted to the $F_2^{\gamma}$ data independently and no attempt was made 
to connect them, whereas in the GS case $F^\gamma_2$ was required to be
the same in LO and NLO at the input scale $Q^2=Q_0^2=3\;GeV^2$: hence they 
are essentially obtained by the same fit to the
data \cite{GS,GSP1,GSP2}. The cross section is less than $2$ pb/GeV at
$p_T^\gamma=2$ GeV, indicating that the full planned luminosity of LEP2 of
$500\;pb^{-1}$ will be necessary to properly measure the cross
section. 

In fig.3(b) we compare the full cross section to the contributions from
fragmentation processes only in LO and NLO. The fragmentation
contributions are most significant in the lower $p_T^\gamma$ region and
have a very significant K-factor, but they are still not large enough in NLO 
to dominate the cross section. So while conclusions drawn from the LO
analysis do not need to be completely revised, we note an important
increase in the fragmentation background in this region when the NLO
corrections are included. 
 
In fig.4(a) we show the $p_T^\gamma$ distribution
$d\sigma/dp_T^\gamma $ evaluated in NLO for the 2-res, 1-res and D
contributions to the cross section. As would be expected, the 2-res
contribution is most significant at low $p_T^\gamma$ but falls off most steeply
with $p_T^\gamma$, whereas the D contribution is least significant at
low $p_T^\gamma$ but increases in importance as $p_T^\gamma$ is increased. The
1-res contribution is intermediate between these two. The physical
explanation for this is simply that all the initial photon's energy goes
into the hard process for the direct contributions whereas only some of it is
available for the hard scattering in the resolved processes, making them
correspondingly less efficient at producing high $p_T^\gamma$ final state
photons. We would thus expect most of the sensitivity to the photon
pdfs to be in the lower $p_T^\gamma$ region.

The last point made above is tested in fig.4(b) where we compare the
full NLO cross section with and without gluon initiated processes included
in the hard scattering cross section. There is a very significant fall
in the cross section when we set $g^\gamma=0$. We thus expect that this
cross section could definitely yield important information on $g^\gamma$
at LEP2, given the planned high luminosities. 

In fig.4(c) we show the cross section at the higher CMS energy of
$\sqrt{s}=500$ GeV. We also show the 2-res, 1-res and D contributions to
the cross section for comparison with LEP2 energies. As might be
expected the cross section is significantly larger at these energies, by
a factor of $3$ at $p_T^\gamma=2$ GeV. The relative
importance of the once-, twice-resolved and direct processes as a
function of $p_T^\gamma$ has not altered significantly. Thus the main
advantage of a machine at this CMS energy would be an increase in the cross
section and thus the possibility of more accurate measurements assuming
the luminosities and anti-tagging conditions are the same as in the LEP2
case.  
  
\section{Conclusions}

This study of prompt photon production in $\gamma \gamma$ collisions 
indicates that the cross section will be measurable at LEP2.  The energy
of the machine turns out to be ideal in the sense that it is high enough 
for there to be a significant contribution from resolved photon
processes, particularly those involving $g^{\gamma}$, and low enough 
that we do not expect a dominant contribution 
from fragmentation processes. Hence a measurement of the cross section 
should yield useful information on $g^{\gamma}$.  The relatively small 
contribution from the fragmentation processes also 
means that the inclusive as opposed to the isolated prompt photon cross 
section will be measurable here, although the need to remove the hadronic 
background from $\pi^0$s will mean that some isolation criterion 
will be imposed. We intend to address the issue of isolation in future
work. 

We concede that the cross section for this interesting process is 
small. However, the cross section for other interesting processes at
LEP2, such as $e^+e^- \to W^+W^-$, are also very small: that it is 
why it imperative for LEP2 to achieve its design luminosity. 

\section{Acknowledgements}
This work supported in part by the U.S.~Department of Energy, Division 
of High Energy Physics, Contract W-31-109-ENG-38.

\pagebreak

\pagebreak

\begin{table}
\begin{center}
\begin{tabular}{|r|r|r|l|}
\hline
   &Initial&Final& \\
Figure&State&State&Subprocess\\
\hline
1(a)&D  &F  &$\gamma \gamma \to q {\overline q}$\\
\hline
1(b)&1-res&F  &$\gamma g \to q {\overline q}$\\
\hline
1(c)&1-res&F  &$\gamma q \to gq $\\
\hline
1(d)&1-res&NF &$\gamma q \to \gamma q $\\
\hline
1(e)&2-res&F  &$q q \to q q \; etc.$\\
\hline
1(f)&2-res&NF &$q {\overline q} \to g \gamma $\\
\hline
1(g)&2-res&NF &$g q \to \gamma q$\\
\hline
\end{tabular}
\caption[Classification of the contributions of fig.(1) according to
initial state (D, 1-res., 2-res.), final state (F or NF), and subprocess
type.] {Classification of the contributions of fig.(1) according to
initial state (D, 1-res., 2-res.), final state (F or NF), and subprocess
type.} 
\end{center} 
\end{table}

\noindent
\begin{flushleft}
{\Large \bf Figure Captions}
\end{flushleft}

\begin{description}
\item[{\bf Fig. 1:}] (a)-(g) Different contributions to the process 
$\gamma \gamma \to \gamma X$, classified in Table 1. (h) Notation used for 
resolved photons and photons produced by fragmentation.
\item[{\bf Fig. 2:}] NLO corrections to the direct process 1(a).   
\item[{\bf Fig. 3:}]   The cross section $d\sigma/dp_T^{\gamma}$ vs
$p_T^\gamma$  integrated over rapidity $|y^\gamma | \leq 2$ 
at CMS energy $\sqrt{s}=180$ GeV. (a) The cross section calculated in LO
(dashed line) and NLO (full line) for the GS photon distributions and
(dotted line) for the GRV NLO photon distributions. (b) The cross section 
showing the full results in LO and NLO and the corresponding
contributions from the fragmentation processes.
\item[{\bf Fig. 4:}](a) $p_T^\gamma $ distribution for the direct,
once-resolved and twice resolved contributions to the cross section in
NLO. (b) The cross section in NLO with and without contributions from
subprocesses initiated by gluons included. (c) Same as (a) but at CMS
energy $\sqrt{s}=500$ GeV, and including the sum of all contributions.

\end{description}
\epsffile{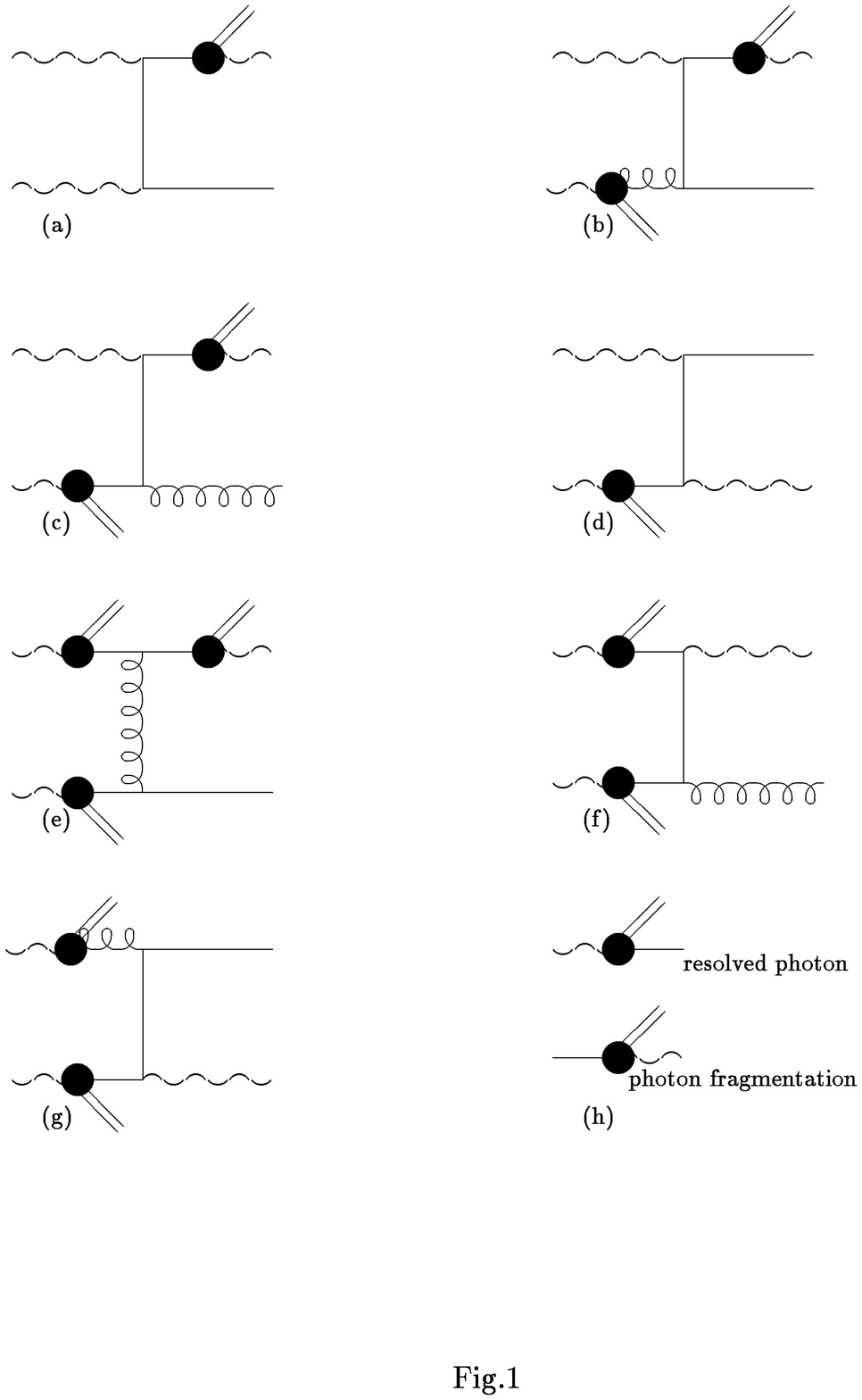}
\epsffile{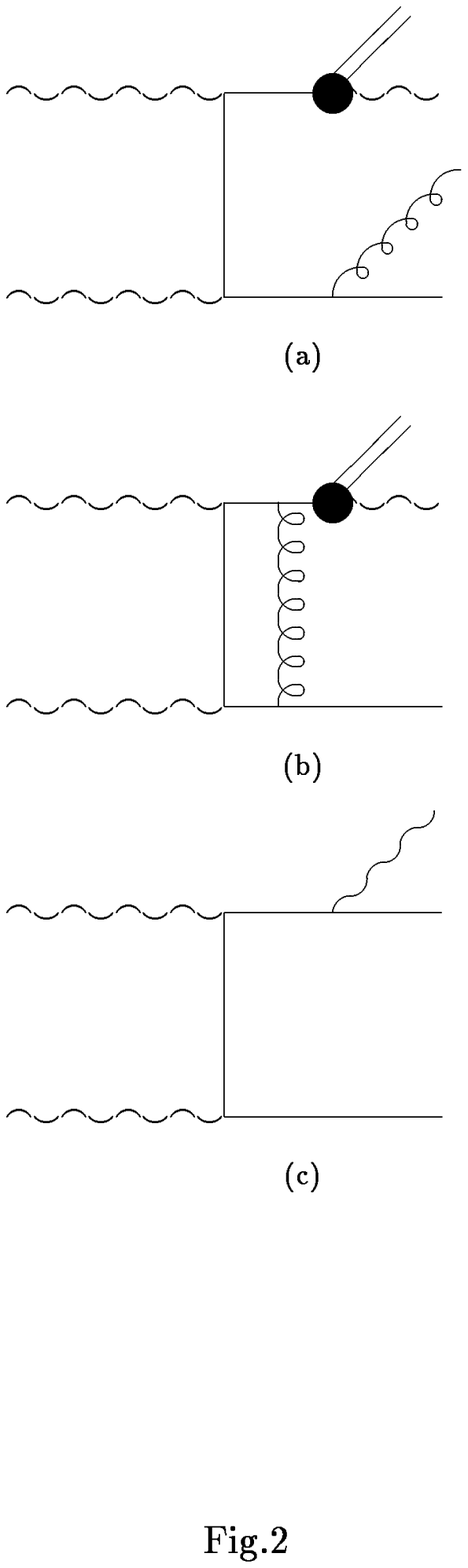}
\epsffile{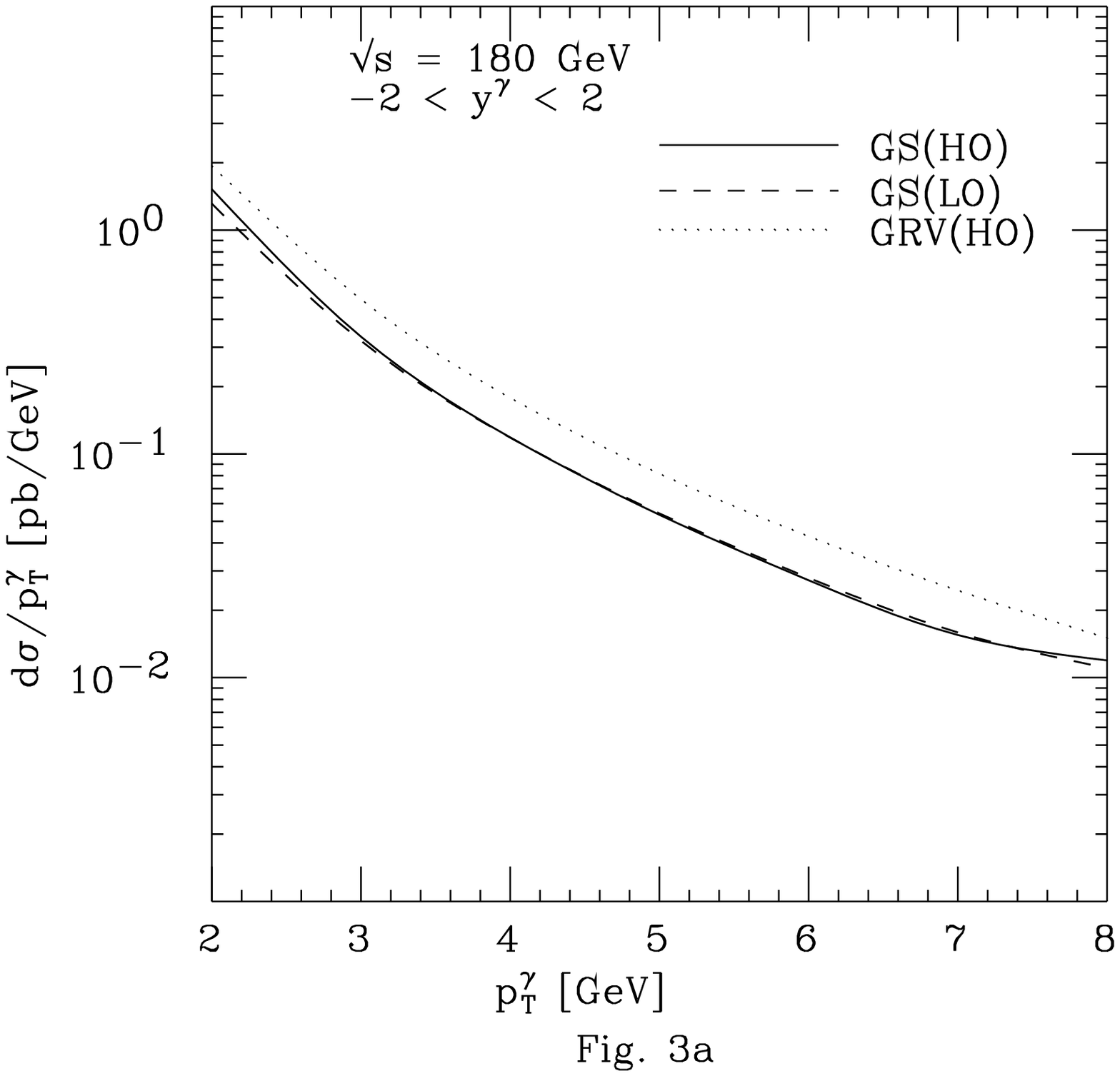}
\epsffile{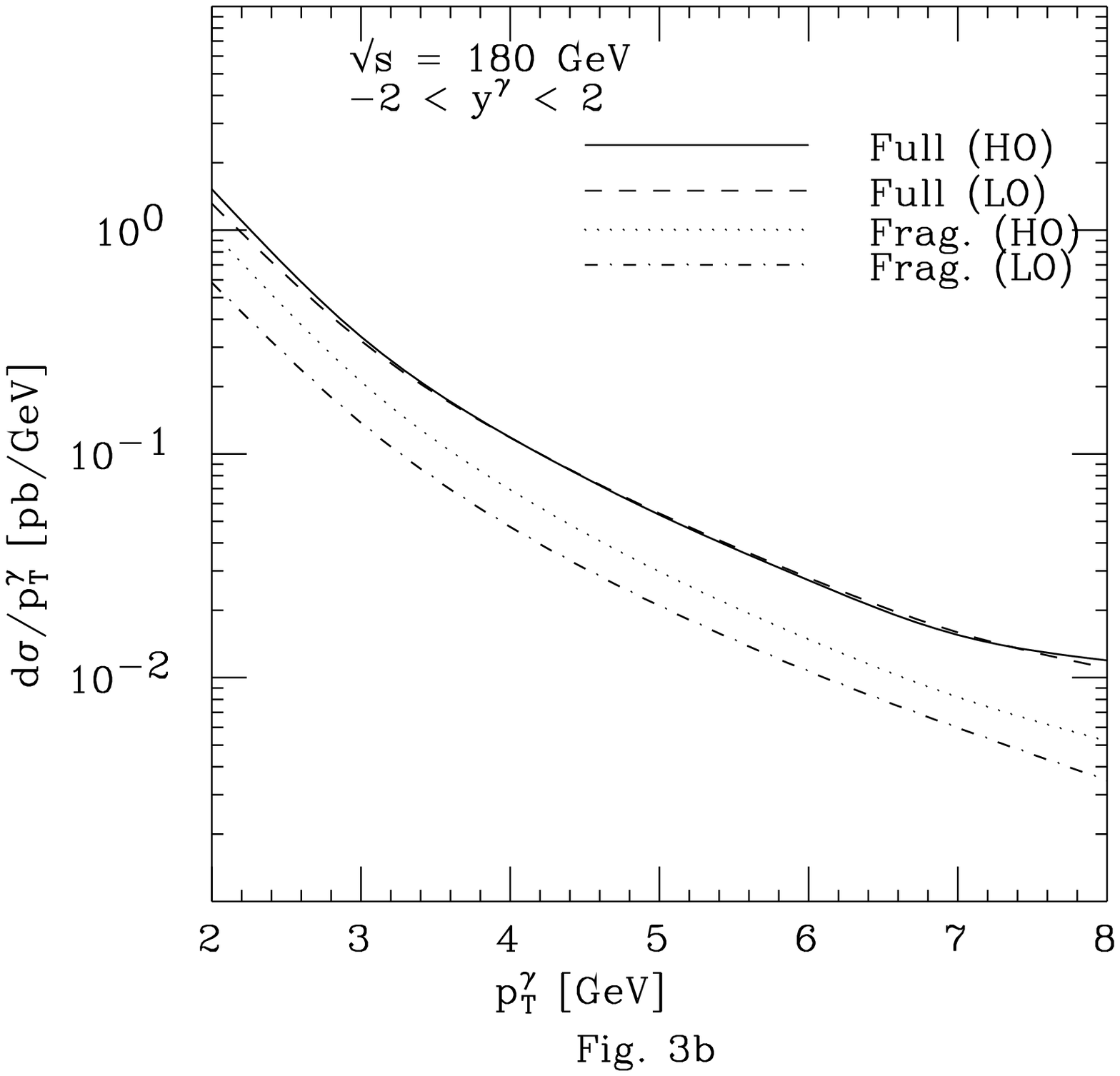}
\epsffile{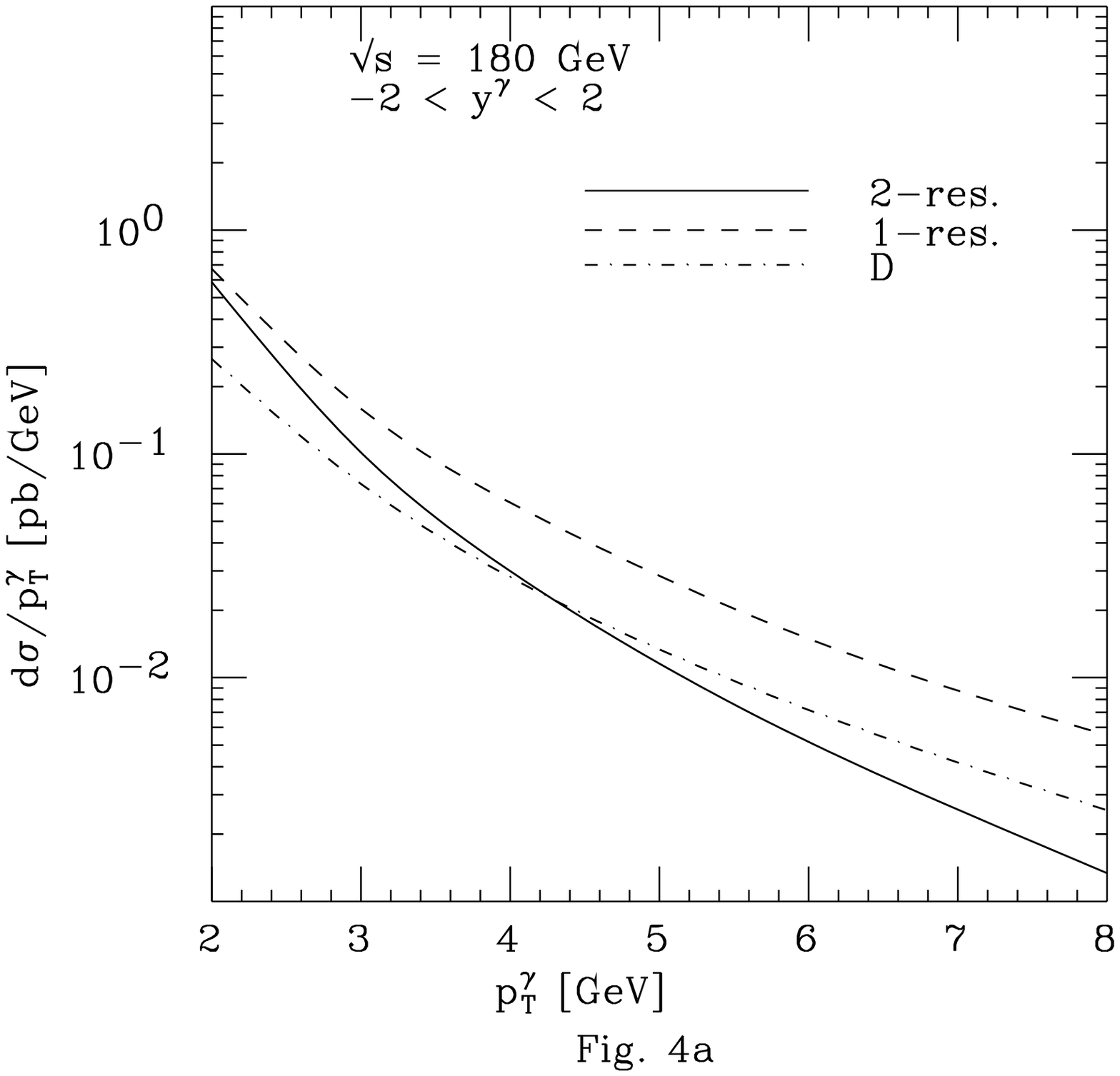}
\epsffile{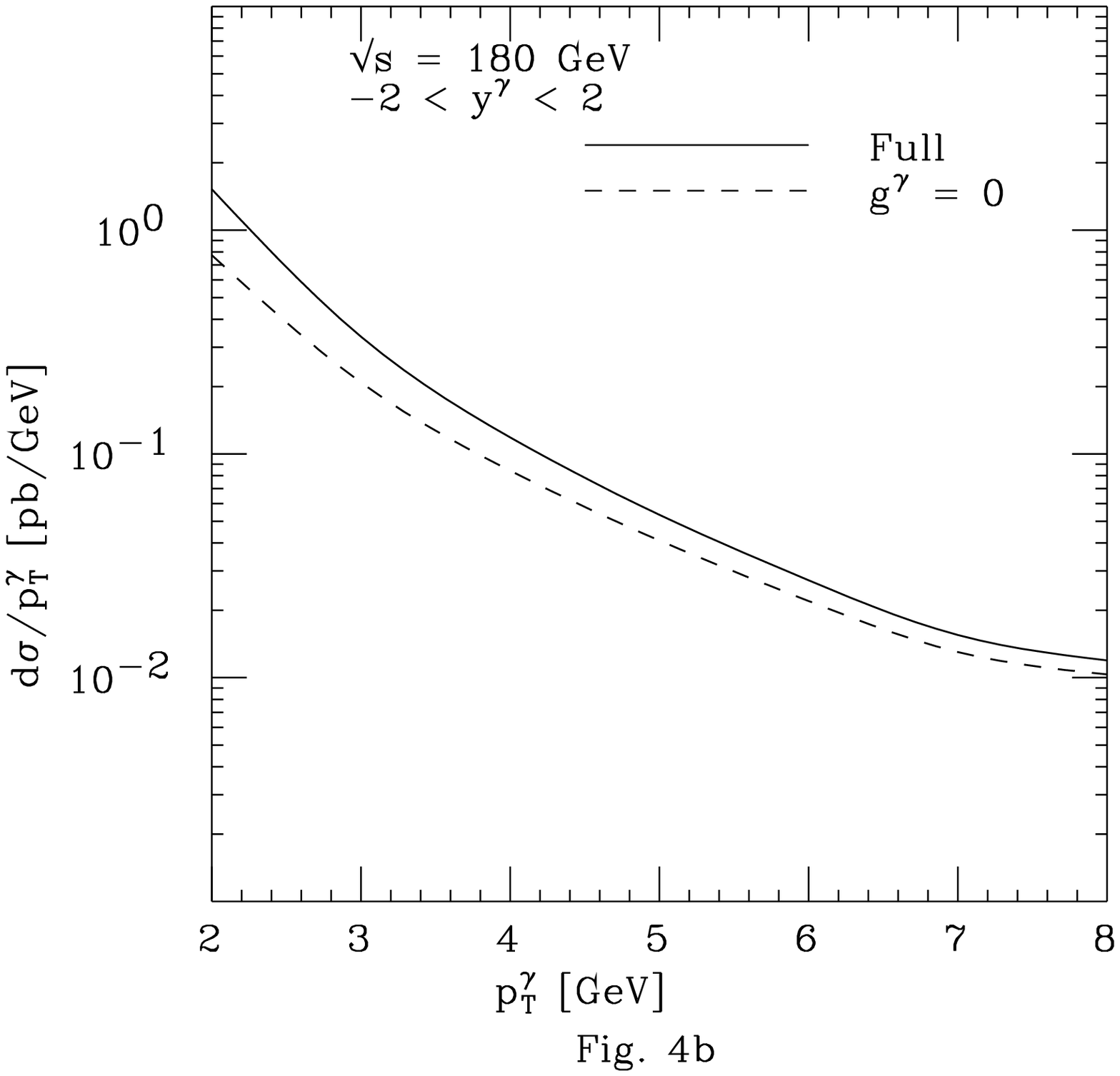}
\epsffile{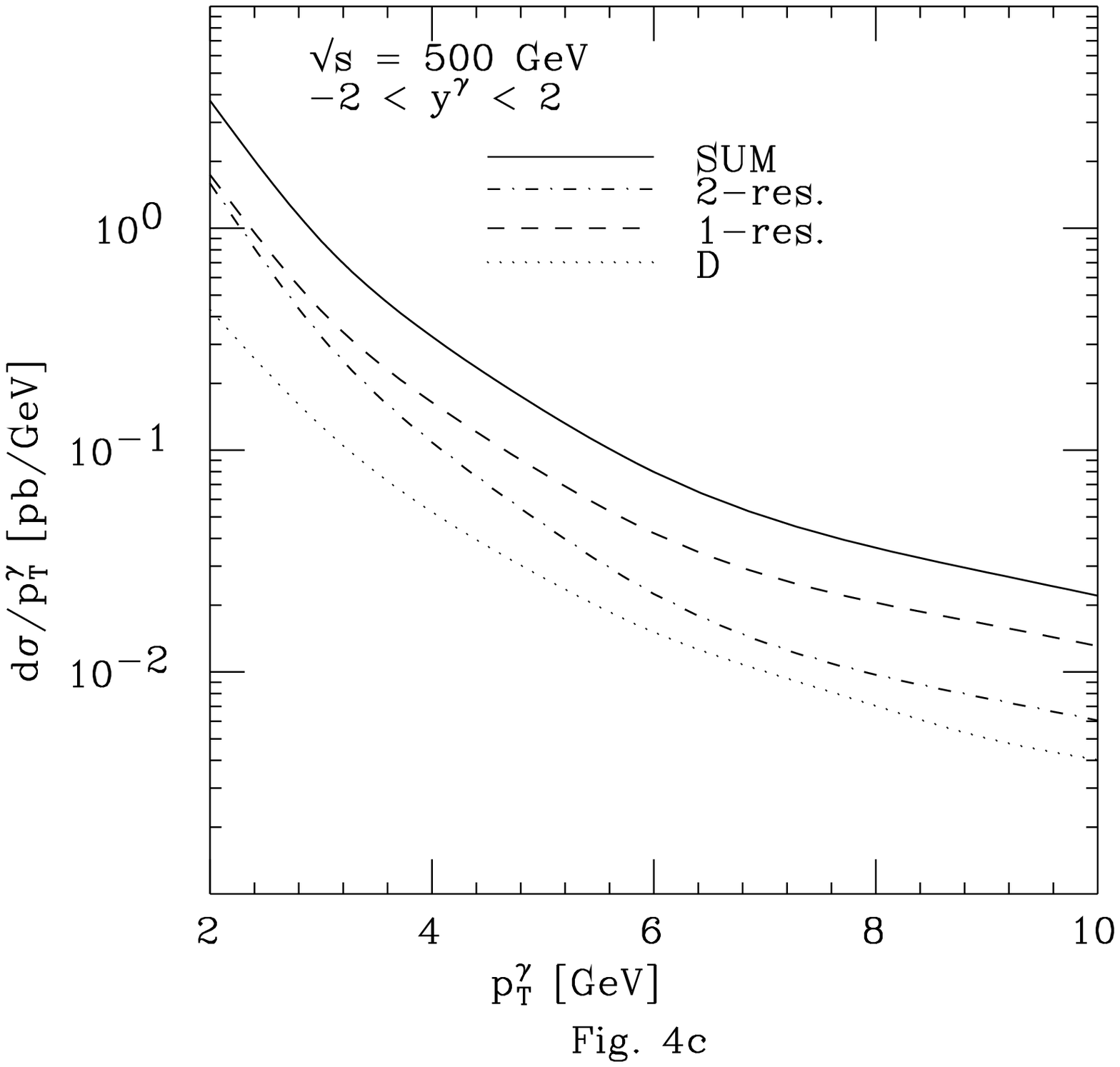}

\end{document}